# Strong Orbital Interaction in π-π Stacking System

Xiao-Xiao Fu[1], Jian-Fu Li[1,2], Rui-Qin Zhang[1*]

**Abstract:**

A simple prototypical model of aromatic π-π stacking system – benzene sandwich dimer is investigated by *ab initio* calculations based on second-order Møller–Plesset perturbation theory (MP2) and Minnesota hybrid functional M06-2X. In succession to our recent exploitation of CH-π complex, strong orbital interaction is also found in the frontier molecular orbitals of benzene sandwich dimer. Similar to interatomic orbital interaction, the intermolecular orbital interaction also forms "bonding" and "anti-bonding" orbitals. The occupied orbitals reveal that a total rise of energy of 7.5 eV occurs when two benzene molecules approach to each other with 3.656 Å. Therein, a large energy rise is caused by repulsive electrostatic interaction between the π electrons of benzene molecules, while a slight decrease is the result of orbital interaction. Our study provides a brand new method to study the π-π stacking system, and throws light on the nature of π-π interaction in an intuitionistic manner.

[1] Department of Physics and Materials Science, City University of Hong Kong, Hong Kong SAR, P.R. China
[2] School of Science, Linyi University, Linyi 276005, P.R. China
[*] E-mail: aprqz@cityu.edu.hk (R.Q.Z.).

The arene-arene interaction has stimulated enormous interest in recent decades, and has been studied extensively both experimentally[1-8] and theoretically[9-28]. The noncovalent intermolecular interaction involving aromatic π molecules is ubiquitous in chemical and biological process and plays an essential role in determining the conformations and properties of molecular assemblies[10,29]. Attractive interaction between π systems controls the base stacking which stabilizes the double helical structure of DNA, and influences the folding of proteins, the self-assembling of supramolecular architectures, and the packing of aromatic molecules in crystalline solids[9]. Also, it is pivotal to protein-ligand recognition and concomitantly to biological or chemical pharmacy[30].

To understand the mechanism of arene-arene interaction, vast researches have been carried out over the years. Besides the solvophobic model, electron donor-acceptor model, and atomic charge model in early stage, a simple electrostatic model was proposed for π-π interaction by Hunter and Sanders in the 1990s[9-11,17]. In this model, the π system was considered as a sandwich of positively charged σ-framework between two negatively charged π-electron clouds, which accounts well for a number of experimental observations. They concluded that the π-σ attraction rather than π-π electronic interaction leads to favourable interactions, where the electrostatic effects determine the geometry of interaction, while van der Waals interactions make the major contribution to the magnitude of the observed interaction. This model provides a qualitative explanation of geometrical preferences for interactions between aromatic molecules, and has been widely recognized. With the improvement of theoretical methods and computational capability, a quantity of researches on benzene dimers based on ab-initio calculations emerged for detailed understanding of the origins, strength, and orientation dependence of π-π interaction[31-38]. These results demonstrated a quantitative relationship between several dominant interactions involved in π-π interactions, and indicated that dispersion associated with electron correlation energy is the major source of attraction in benzene dimers, overcoming the repulsive electronic interaction. However, most of the calculations were concentrated on pursuing the high-accuracy of geometries and energies, and attempted to give explanation of interaction components via complicated calculations and inexplicable analyses. The intuitive quantum-mechanical level nature of mechanism of the π-π interaction of is rarely found.

As is known, orbital interaction theory provides a comprehensive model for examining the structures and kinetic and thermodynamic stability of molecules, facilitating our understanding of the fundamental processes in chemistry. It has been known that orbital interactions exist not only between atoms but also within special organic molecules through space and through bond. Our recent work demonstrated an obvious intermolecular orbital interaction between benzene and methane in complex, which is considered as a representative of weak CH-π interaction system[39]. Herein, taking the dimer of two face-to-face benzene molecules with sandwich configuration as the prototype, the π-π interaction is systematically investigated based on ab-initio calculations. Focusing on molecular orbital analysis, we studied the density of states (DOS), absorption properties, and charge distribution of the dimer. Interestingly, we also find a conspicuous intermolecular orbital interaction between benzene molecules in the dimer, which reveals the nature of π-π interaction from an intuitionistic viewpoint, distinctive from the previous studies.

## Results

The configuration of benzene sandwich dimer was built referring to previous model[40], where the two benzene molecules are located face-to-face. The basis set superposition error (BSSE) was taken into account for interaction energy by counterpoise correction. The BSSE-corrected interaction energies at different dimer interaction distance calculated by MP2/cc-pVTZ are shown in Figure 1, which presents a binding energy of -2.508 kcal mol$^{-1}$ at the equilibrium distance 3.8 Å. The result compares well with previous results, -2.557 kcal mol$^{-1}$ at MP2/aug-cc-pVDZ level, with $R$=3.9 Å taken from MP2/DZ+2P geometry calculation[35]. Further relaxation derived the BSSE-corrected intermolecular interaction energy of -2.455 kcal mol$^{-1}$ with the intermolecular distance $R$ equaling to 3.656 Å.

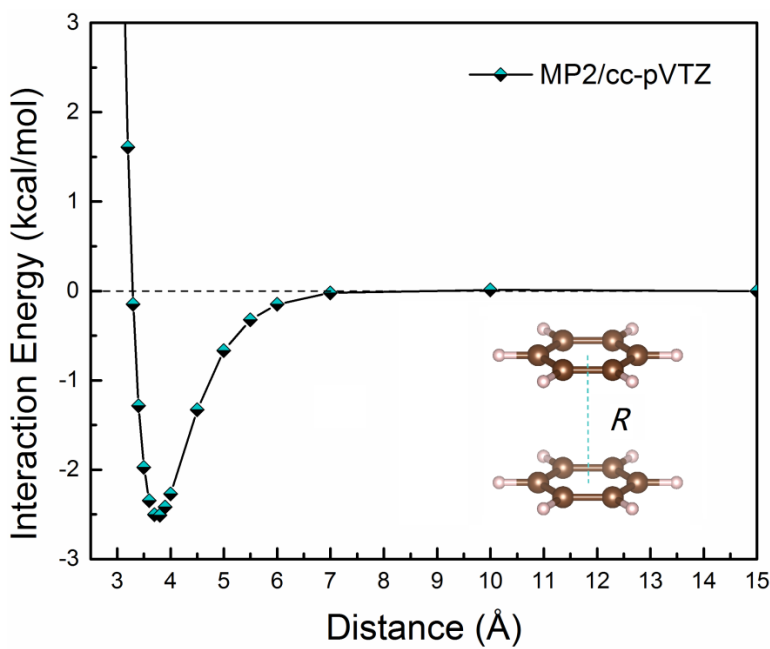

Figure 1 The BSSE-corrected intermolecular interaction energy of benzene sandwich dimer calculated by MP2/cc-pVTZ. The insert is the schematic structure of sandwich configuration, and $R$ is the interaction distance which is measured from the centroids of two benzene molecules.

**Orbital distribution.** The spatial distributions of the frontier molecular orbitals together with their corresponding energy levels of the benzene sandwich dimer and the benzene monomers are plotted in Figure 2 (The frontier LUMOs are shown in Figure S1). All the orbitals of the dimer including occupied and unoccupied orbitals are distributed on both benzene rings equally, and the composition of orbitals from either part is exactly 50%. For benzene monomer in Figure 2, we consider the frontier five occupied molecular orbitals for analysis, where two degenerate states at both -9.15 eV and -13.37 eV, and one nondegenerate state at -13.68 eV; they are HOMO ~ HOMO-1, HOMO-2 ~ HOMO-3, and HOMO-4, respectively (see Table 1). The orbitals at the same energy level from two benzene molecules in dimer form a couple of bonding and anti-bonding orbitals of the dimer, similar to interatomic orbital interaction. For example, HOMO and HOMO-2 of the dimer is a pair of anti-bonding and bonding orbitals composed of HOMO of two benzene molecules, and HOMO-1 and HOMO-3 of the dimer correspond to HOMO-1 of benzene molecules. Bonding orbitals are more delocalized with lower energy levels in comparison with anti-bonding orbitals. Examining the bonding orbitals of dimer, we can find that under the certain isosurface value of ±0.02, some of them consist of two separated components from benzene monomers, and the orbital deformation is negligible, such as HOMO-6 and HOMO-7; while for the others, like HOMO-2, HOMO-3 and HOMO-9, the orbitals located not only on the two individual benzene parts, but also the intermediate regions between them. A notable deformation of orbitals occurs under π-π interaction, very similar to that yielding a covalent bonding. Moreover, a large energy difference is found to be accompanied with remarkable orbital deformation. In fact, all the molecular orbitals at the same energy levels from two benzene molecules overlap with each other, forming new bonding and anti-bonding orbitals of the dimer. However, only a slight deformation and small energy difference exist for most pairs of bonding and antibonding orbitals of the dimer (Figure S2 is for demonstration with lower isosurface value, and Figure S3 shows the energy levels of all occupied orbitals for benzene sandwich dimer and the benzene monomer).

In addition, it should be noted that the energy levels of HOMO, HOMO-1 and HOMO-4 of benzene monomer are located in intermediate regions between those anti-bonding and bonding orbitals they composed. But, for HOMO-2 and HOMO-3 of benzene monomer, their energy levels are slightly lower than the corresponding anti-bonding and bonding orbitals of dimer. This is caused by the slight shift of energy levels of the dimer due to electronic interaction when two π-electron systems approach to each other, and also the weak deformation of the bonding orbitals. Since orbitals provide the most fundamental properties of electrons of system, the strong orbital interaction of the benzene sandwich dimer should be convictive evidence supporting the relatively intense interaction of π-π stacking systems.

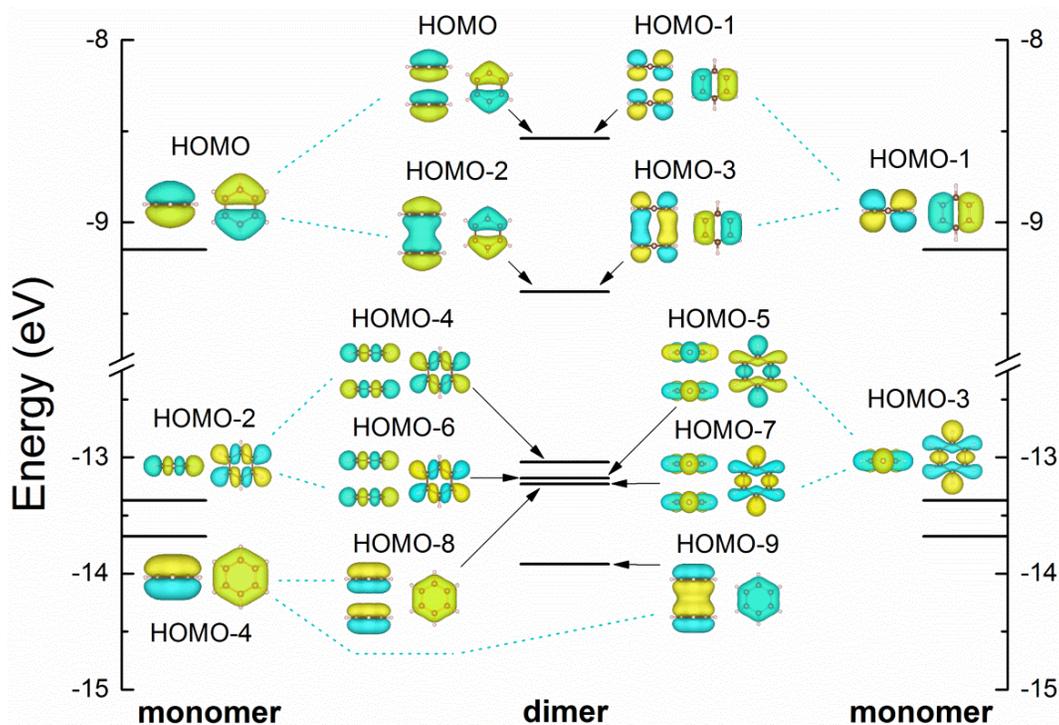

Figure 2 The frontier occupied molecular orbitals (side view and top view) together with energy levels of the benzene sandwich dimer and the benzene monomer based on MP2/cc-pVTZ level calculation. The isosurface value is ±0.02, where the yellow and green colors denote positive and negative values, respectively. The green dashed line denotes the relation between the original orbitals of benzene monomer and their corresponding bonding and anti-bonding orbitals of the dimer.

Table 1 The energy values (in eV) of HOMOs for both benzene dimer and monomer shown in Figure 2.

| MOs of dimer | Energy | MOs of monomer | Energy |
|---|---|---|---|
| HOMO, HOMO-1 | -8.54 | HOMO, HOMO-1 | -9.15 |
| HOMO-2, HOMO-3 | -9.38 | | |
| HOMO-4 | -13.04 | HOMO-2, HOMO-3 | -13.37 |
| HOMO-5, HOMO-6 | -13.18 | | |
| HOMO-7, HOMO-8 | -13.23 | HOMO-4 | -13.68 |
| HOMO-9 | -13.92 | | |

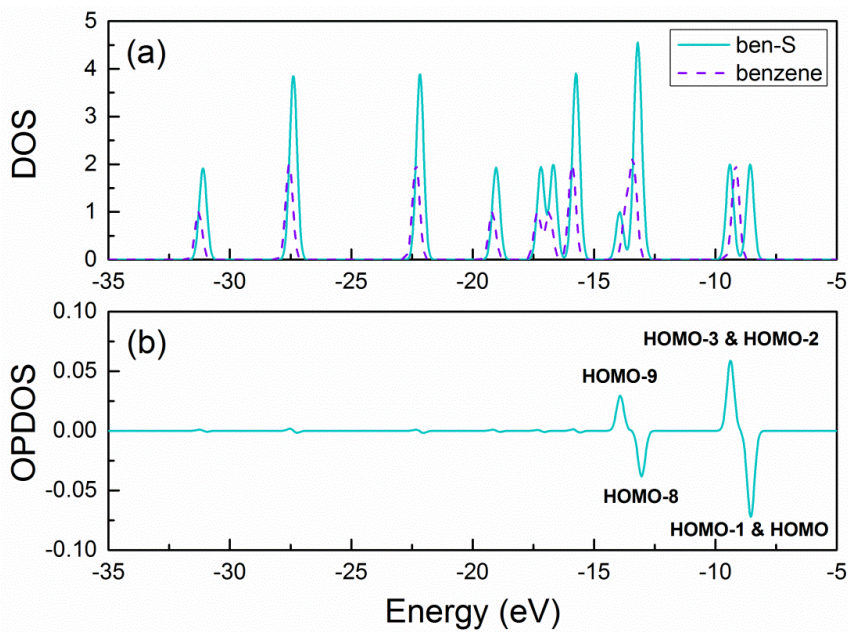

Figure 3 DOS and OPDOS of benzene sandwich dimer (ben-S) within shallow occupied energy level region (-35.0 eV ~ -5.0 eV) calculated at MP2/cc-pVTZ level of theory. The green solid lines are DOS and OPDOS for ben-S, and the purple dashed line denotes the DOS of benzene monomer. The two benzene molecules in ben-S dimer contribute equally to DOS.

**Density of states.** To elucidate the variation of molecular orbitals with respect to energy levels before and after the dimer formation, the DOS of isolated benzene and sandwich dimer were calculated using MP2/cc-pVTZ (Figure 3(a)). Since the PDOS of benzene molecules of sandwich dimer are identical and overlap with each other absolutely, their contributions to total DOS of the dimer are equal. The results indicate that the formation of the sandwich dimer causes not only the orbital superposition of two benzene molecules, but also the energy shift of both parts. Most of the energy levels of the dimer are left-shifted (about 0.15 eV) compared with those of isolated benzene molecule due to π-electron interaction between two parts. Relative to the isolated benzene, the energy levels of the dimer in the ranges of (-10 eV, -7.5eV) and (-15.0 eV, -12.5 eV) are markedly changed, which results from the deformation of HOMOs, mainly HOMO and HOMO-2, combined with HOMO-1 and HOMO-3, and also HOMO-8 and HOMO-9. On the contrary, in the region where only a slight shift of energy level occurs, the orbitals contain no obvious overlap between two benzene molecules.

OPDOS, the overlap projected density of states, also referred to as crystal orbital overlap population (COOP) in the literature, offers a measure of the overlap strength and an energy resolved quantity where positive peak indicates bonding state and negative means anti-bonding state. Based on this concept, the OPDOS between two benzene molecules in the dimer (Figure 3(b)) was calculated via GaussSum program[41] at MP2/cc-pVTZ level of theory. The HOMO-9, HOMO-8, HOMO-3 and HOMO-2 (degenerate), and HOMO-1 and HOMO (degenerate) show strong orbital overlap between two benzene molecules, and their corresponding strengths are 0.029, 0.038, 0.059 and 0.072. The overlap strength of each orbital is about 20% of that between carbon and hydrogen atoms in methane (the strength is $0.15^{39}$). OPDOS provides a quantitative proof for the strong orbital overlap in the benzene dimer.

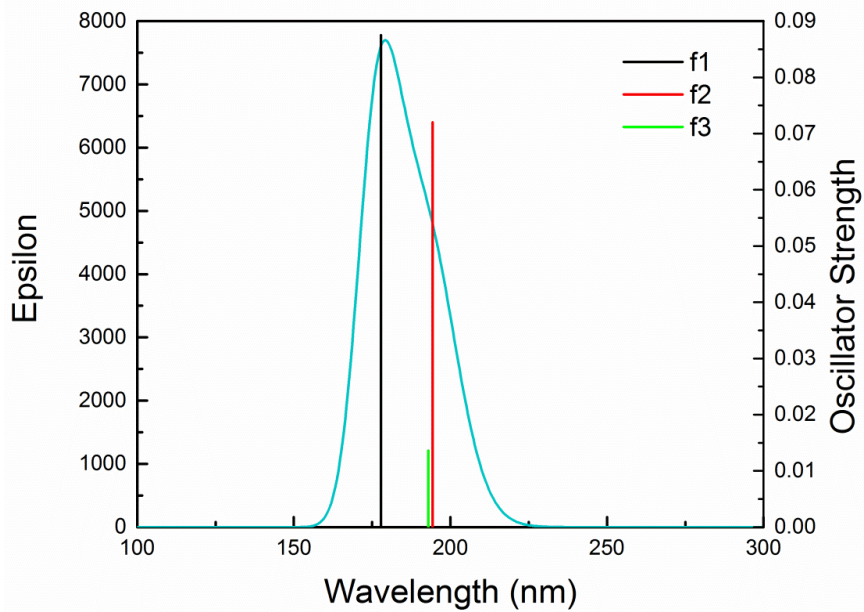

Figure 4 Absorption spectra for benzene sandwich dimer calculated at M06-2X/6-311++G** level of theory. The f1 ~ f3 denote three absorption peaks with maximum oscillator strength.

**Absorption properties.** As reported by Hunter's et al[9], π-π interaction could rarely cause a distortion of UV-visible spectra of two chromophores in the experiment, so they considered that the two interacting π-systems do not distort each other's molecular orbitals. The absorption properties of benzene sandwich dimer are shown in Figure 4, which is calculated at M06-2X/6-311++G** level of theory. From our results, there exists a prominent absorption of benzene sandwich system with the spectra range from 225nm to 150 nm in the ultraviolet and far-ultraviolet region, which is in agreement with the statement of spectra above. However, for the orbitals, our results reveal a strong orbital deformation in benzene sandwich dimer. By checking the transition states with the maximum oscillator strength, we found that transition happens mainly from the frontier HOMO ~ HOMO-3 to LUMOs. From previous discussion, we have known that the orbital deformation takes place in frontier π or π* orbital and leads to a large difference between energy levels, which should facilitate a high energy absorption. Thus, the absorption of ultraviolet and far-ultraviolet spectra is consistent with the strong deformation of orbitals for the π-interaction system.

## Discussion

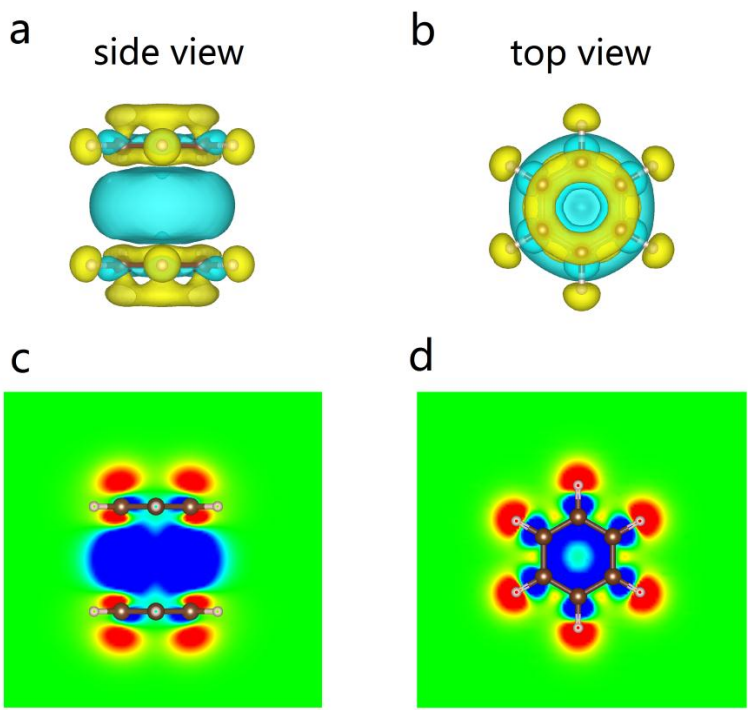

Figure 5 Electron density difference for benzene dimer calculated at MP2/cc-pVTZ level: a and b, isosurface plotting of side and top view, respectively, where the blue region represents the electron donor, and the yellow region represents the electron acceptor; c and d, longitudinal section of the dimer and the cross section containing benzene, respectively, where the blue region represents the electron donor, and the red region represents the electron acceptor. The isosurface value is ±0.00015.

The electron density difference between the dimer and the isolated molecules demonstrates the consequence of intermolecular interaction between two benzene molecules, as shown in Figure 5. The intermolecular interaction between occupied orbitals generates new molecular orbitals, leading to rearrangement of the electronic distribution. With the isosurface value equal to ±0.00015, obvious charge transfer occurs in the region between and surrounding two molecules. Relative to isolated molecules, both the intermediate region of dimer and the skeleton of benzene molecules are absent of electron, while the region under and above the benzene skeleton, together with the hydrogen atoms obtain electrons. The multipole moments are modulated due to intermolecular interaction, where the magnitude of quadrupole moment is moderately enlarged, the octapole moment emerges, and the hexadecapole moment is largely changed (see Table S1). With respect to Mulliken atomic charges, the absolute values of charges distributed on each carbon and hydrogen atom in dimer are slightly increased by 0.7% compared to isolated benzene molecule. Besides, the electron density difference indicates a significant electrostatic repulsive interaction between two molecules in dimer. Since benzene is an electron-rich molecule, the proximity of two benzene molecules in sandwich configuration will lead to repulsion of π-electron clouds. This is considered to be the dominant interaction between π-π stacking systems and reaches common agreement.

By subtracting all the occupied energy levels of the dimer and those of two isolated benzene molecules, we found that a total energy rise of 7.5 eV occurs when two benzene molecules approach to form a sandwich dimer, and the majority of that rise should be attributed to electrostatic interaction (see Figure S3). A total energy split of 3.38 eV is the result of the occupation of the bonding and anti-bonding orbitals induced by the overlap of intermolecular orbitals. The stronger orbital interaction brings the larger energy differences between the anti-bonding and bonding orbitals, together with smaller energy rise compared with those weaker interaction orbitals. In details, the occupation of strong interaction orbitals (HOMO ~ HOMO-3, HOMO-8 and HOMO-9) results in a total energy rise of 0.97 eV, averagely 0.1617 eV per orbital, while occupation of the rest occupied orbitals leads to an energy enhancement of 6.53 eV totally and 0.1814 eV averagely. The energy rise caused by the strong interaction orbitals is lower than the weak interaction orbitals, despite of the fact that the frontier orbitals at shallow energy levels should be more sensitive to electrostatic interaction than those orbitals at deep energy levels. Therefore, the orbital interaction should be in favor of total energy decrease and the stabilization of dimer. Although sandwich configuration is considered to have weaker intermolecular interaction and also to be less preferable in liquid or solid, it enables a larger contact surface which is expected to experience greater gains in London-dispersion-induced stabilization.[33,34] That means the London dispersion plays a dominant role to stabilize the sandwich dimer, overcoming the electronic repulsion. In spite of the fact that the density functional theory (DFT) we employed in this work relies on local approximation of density and is incapable of accurately describing dispersion force, we found that the factors advantageous to dispersion are in accord with those to orbital interaction. Thus, we supposed that the orbital interaction should be associated with dispersion interaction; at least it should be an average or a part of expression of dispersion.

Moreover, comparing the features of HOMO ~ HOMO-4 of benzene monomer and their corresponding bonding and anti-bonding orbitals of dimer, we can find that those orbitals with delocalized spatial distributions, such as HOMO, HOMO-1 and HOMO-4, are more likely to form deformed bonding orbitals, HOMO-2, HOMO-3, and HOMO-9, respectively. Although the other occupied orbitals of two benzene molecules in sandwich dimer are located at the same energy level, only those with π or π* orbital can result in a remarkable orbital interaction. In our recent work on CH-π system, we found that the orbital deformation is determined by two conditions: one is the delocalization of π or π* orbitals and the other one is the closeness in energy between the π or π* orbital of benzene and the σ orbital of methane[39]. Hereon, in π-π interaction system, further conclusion can be drawn: As the orbitals of two molecules in dimer are located at the same energy level, all the orbitals present the orbital superposition, but, only those π or π* orbitals with substantial delocalization can lead to strong orbital interaction.

## Conclusion

The systematic analysis of electronic properties of benzene sandwich dimer from ab-initial calculation based on MP2/cc-pVTZ and M06-2X/6-311++G** level, including molecular orbital, DOS, OPDOS, absorption spectra, and electron density difference, reveals a strong orbital interaction is found in π-π interaction system, which enrich the existed understanding of the nature of π-π interaction. DOS and OPDOS present consistent results with molecular orbitals. The absorption spectra of the system present the ultraviolet and far-ultraviolet absorption, giving another evidence for the intense orbital interaction in frontier occupied orbitals. The electron density provides an overall consequence of π-π interaction.

## Method

**Computational methods.** The *ab initio* calculations of structural relaxations and electronic properties of the benzene sandwich dimer were carried out with the Gaussian 09 package[42]. The second-order Møller–Plesset perturbation theory (MP2) combined with cc-pVTZ basis set was used to explore the configuration and interaction energy in the ground state. MP2 is the most commonly used method to describe intermolecular interactions due to its inclusion of electron correlation for capturing dispersion interaction, which approximately accounts for the uncoupled and two-body electron correlations. The basis set superposition error (BSSE) was taken into account for interaction energy by counterpoise correction. Another method/basis set M06-2X/6-311++G**, was chosen to repeat all the calculation for confirmation, and to obtain the absorption property. Both MP2/cc-pVTZ and M06-2X/6-311++G** are reliable for weak interaction system according to previous calculation[33,39]. Single point calculations were used to show the impact of intermolecular interaction on orbitals between two benzene monomers.

**Acknowledgements**

The work described in this paper was supported by a grant from the Research Grants Council of Hong Kong SAR (Project No. CityU 103913). This work utilized the High Performance Computer Cluster managed by the College of Science and Engineering of City University of Hong Kong, and the High Performance Cluster Computing Centre of Hong Kong Baptist University, which receive funding from the Research Grant Council, the University Grant Committee of the HKSAR, and Hong Kong Baptist University.

**Author contributions**

X.-X. Fu carried out the theoretical calculations and wrote the manuscript. J.-F. Li offered a great help in calculation and discussion. R.-Q. Zhang was responsible for the overall direction.

**Additional information**

Supplementary Information accompanies this paper at http://www.nature.com/naturecommunications

**Competing financial interests**: The authors declare no competing financial interests.

# Strong Orbital Interaction in pi-pi Stacking System

Xiao-Xiao Fu[1], Jian-Fu Li[1,2], Rui-Qin Zhang[1*]

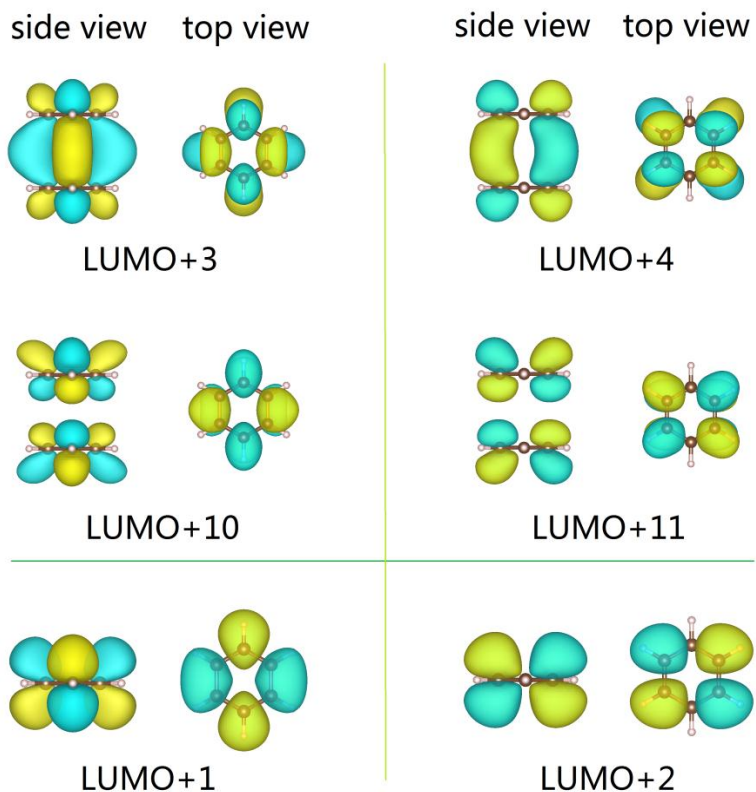

Figure S1 The frontier unoccupied molecular orbitals (LUMO+3, LUMO+4, LUMO+10 and LUMO+11) of benzene sandwich dimer and the original orbital in monomer, based on MP2/cc-pVTZ level calculation. The isosurface value is ±0.02, where the yellow and green colors denote positive and negative values, respectively.

In Figure S1, the LUMO+1 and LUMO+2 of benzene monomer are degenerate molecular orbitals. The LUMO+1 of both benzene molecules in dimer form the bonding orbital LUMO+3 and antibonding orbital LUMO+10 of dimer. And the LUMO+2 of benzene monomer corresponds to LUMO+4 (bonding orbital) and LUMO+11 (anti-bonding orbital) of dimer. Therein, LUMO+3 and LUMO+4 are degenerate, and the same to LUMO+10 and LUMO+11.

[1] Department of Physics and Materials Science, City University of Hong Kong, Hong Kong SAR, P.R. China
[2] School of Science, Linyi University, Linyi 276005, P.R. China
* E-mail: aprqz@cityu.edu.hk (R.Q.Z.).

With isosurface value of ±0.02, no obvious deformation of HOMO-4 ~ HOMO-7 is found. Decreasing the isosurface value to ±0.005, a certain deformation of those orbitals occurs, as shown in Figure S2 That means the deformation of HOMO-4 ~ HOMO-7 can be negligible compared to HOMO ~ HOMO-3.

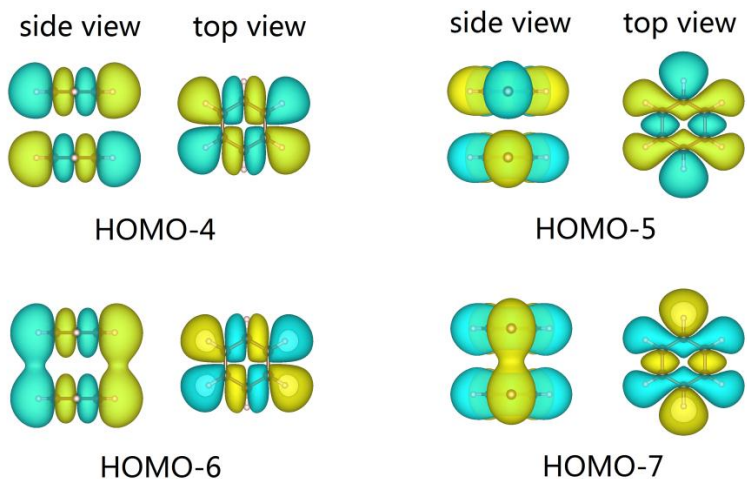

Figure S2 The frontier molecular orbitals (HOMO-4 ~ HOMO-7) of benzene sandwich dimer based on MP2/cc-pVTZ level calculation. The isosurface value is ±0.005, where the yellow and green colors denote positive and negative values, respectively.

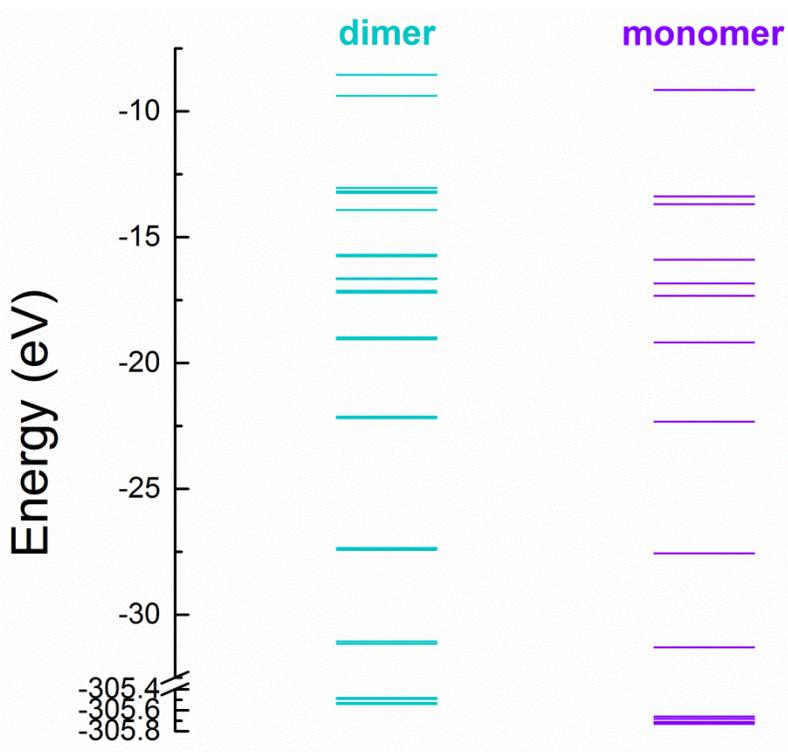

Figure S3 The energy levels of occupied orbitals for benzene sandwich dimer and the benzene monomer based on MP2/cc-pVTZ level calculation.

From Figure 3S, we can find that most of the new bonding and anti-bonding orbitals of dimer are very close in energy, but the energy shift is visible, which is due to electronic interaction of π-electron of benzene molecules.

Table S1 Multipole moments of the benzene monomer and sandwich dimer (D: Debye) based on MP2/cc-pVTZ level calculation.

| | Quadrupole moment (in D·Å) | | | | | |
|---|---|---|---|---|---|---|
| | XX | YY | ZZ | XY | XZ | YZ |
| Monomer | -31.8374 | -31.8387 | -40.6526 | 0.0000 | 0.0000 | 0.0000 |
| Dimer | -64.2874 | -64.2900 | -83.1686 | 0.0000 | 0.0000 | 0.0000 |
| | Octapole moment (in D·Å$^2$) | | | | | |
| | XXX | YYY | ZZZ | XYY | XXY | |
| Monomer | 0.0000 | 0.0000 | 0.0000 | 0.0000 | 0.0000 | |
| Dimer | 0.0000 | 0.0000 | -456.0446 | 0.0000 | 0.0000 | |
| | XXZ | XZZ | YZZ | YYZ | XYZ | |
| Monomer | 0.0000 | 0.0000 | 0.0000 | 0.0000 | 0.0000 | |
| Dimer | -117.5040 | 0.0000 | 0.0000 | -117.5087 | 0.0000 | |
| | Hexadecapole moment (in D·Å$^3$) | | | | | |
| | XXXX | YYYY | ZZZZ | XXXY | XXXZ | |
| Monomer | -270.1477 | -270.1646 | -48.5605 | 0.0000 | 0.0000 | |
| Dimer | -545.8872 | -545.9216 | -3410.1107 | 0.0000 | 0.0001 | |
| | YYYX | YYYZ | ZZZX | ZZZY | XXYY | |
| Monomer | 0.0000 | 0.0000 | 0.0000 | 0.0000 | -90.0534 | |
| Dimer | 0.0000 | 0.0000 | 0.0000 | 0.0000 | -181.9708 | |
| | XXZZ | YYZZ | XXYZ | YYXZ | ZZXY | |
| Monomer | -64.4215 | -64.4228 | 0.0000 | 0.0000 | 0.0000 | |
| Dimer | -561.1945 | -561.2139 | 0.0001 | 0.0000 | 0.0000 | |